\documentstyle{article}

\begin{document}

\def\lg{\mbox{$\cal G$}}

\def\hf{\mbox{$\bar{H}$}}
\def\hs{\mbox{$\tilde{H}$}}
\def\hl{\mbox{$H^{\omega}$}}
\def\hv{\mbox{$\hat{H}$}}

\def\mcolon{{\vcenter{\hbox{\scriptsize\boldmath $:$}}}}
\def\ldef{\mathrel{\hbox{$\mcolon \mkern0.75mu \mathord{=}$}}}
\def\rdef{\mathrel{\hbox{$\mathord{=} \mkern1mu \mcolon$}}}

\newcommand{\Reell}
  {\mbox{\bf I$\!$R}
  }

\newcommand{\leftsemi}
  {\;\!\times\!\!\!\!\!\!\supset\;\!\!
  }

\newcommand{\rightsemi}
  {\;\times\!\!\!\!\!\!\!\subset
  }

\newcommand{\deflabel}[1]{\bf #1\hfill}

\newenvironment{deflist}[1]{
   \begin{list}{}
      {
      \settowidth{\labelwidth}{\bf #1}
      \setlength{\leftmargin}{\labelwidth}
      \addtolength{\leftmargin}{\labelsep}
      \renewcommand{\makelabel}{\deflabel}
      }
   }{
   \end{list}
    }

\title{$TG$--EQUIVARIANCE OF CONNECTIONS AND  GAUGE TRANSFORMATIONS}
\author{H. M. Bosselmann\\
        N.A. Papadopoulos\\
        Institut f\"ur Physik\\
        Johannes Gutenberg-Universit\"at\\
        D-55099 Mainz, Germany}

\date{December 6, 1994}

\maketitle

\begin{abstract}
We consider the notion of a connection on a principal bundle $P(M,G)$ 
from the point of view of the action of the tangential group $TG$. This 
leads to a new definition of a connection and allows to interpret gauge 
transformations as effect of the $TG$ action. 
\end{abstract}

\section{Introduction}
The central role the gauge potential plays in the theories of 
fundamental interactions of nature always justifies new investigations 
concerning its nature. 
If we take into account gauge transformations it becomes quite apparent 
that gauge potential is something special compared to any other particle 
field.
It is not a field which transforms 
itself like a tensor field. It is something more complicated than a 
tensor field with more involved and deeper lying properties. 

A gauge potential is carrier of a geometric structure in the sense of 
the metric structure in the general relativity, but in a more general 
sense.
Because of its complexities, its deepness and its importance, there 
exist several formulations of it [1, 2]. In this paper, we would like to 
present a new property of the gauge potential: the $TG$--equivariance of 
a connection on a principal bundle $P(M,G)$ with $M$ the space-time 
(basis manifold) and $G$ the stucture group. The $G$--equivariance 
property of a connection is of course well known. Here we mean, as we 
shall explain, the equivariance respectively to the bigger group, the 
tangential group $TG$, of the structure group $G$. It is also important 
to note that the $G$--equivariance property is shared by the usual tensor 
fields, considered as vector-valued functions on a principal bundle 
$P(M,G)$. But only the connection has in addition the $TG$--equivariance
property. 

The motivation for this paper came from considerations within the 
$G$--theory framework and applications of it in investigating the 
structure of the space of reducible connections for Yang-Mills theories 
[3], anomalies [4, 5] and recently within a non commutative geometrical 
approach to the standard model [6]. The $G$--theory point of view 
emphasizes particularly the action of a group and all what follows from 
it on a  relevant space of physical objects. 

Our starting point is as already mentioned a principal bundle $P(M,G)$ 
with a connection $\omega$ on it and the structure group $G$ acting by 
definition freely on $P$ from the right. Going now to the tangential 
space $TP$ and $TG$ of $P$ and $G$ respectively, it is important to 
realize that $TG$ may be considered as a Lie group and that it acts 
freely on $TP$. The multiplication law $TG \times TG \rightarrow TG$ and 
the action $TP \times TG \rightarrow TP$, are to be defined below. This 
leads to the fact that the tangential space $TP$ is also a principal 
bundle with basis manifold $TM$ and structure group $TG$. We now 
consider the space $TP/G$ as a fiber bundle (non-vector bundle) with 
typical fibre the Lie algebra $\lg$ of $G$. We show that $TP/G$ is 
isomorphic to the associated fiber bundle of $TP : TP \times_{TG} \lg$ 
where the action of $TG$ on $\lg$ is deduced from the multiplication law 
in $TG$. We give a new definition of a connection on $P$: a 
connection on $P(M,G)$ is a linear section $\hs$ on the fiber bundle $TP 
\times_{TG} \lg$ 
\begin{eqnarray}\label{eqn1}
        \hs : TM &\rightarrow& TP \times_{TG} \lg \quad.
\end{eqnarray}
It can be shown that $\omega$ (from the usual definition of a 
connection) and $\hs$, as above, are equivalent. The section $\hs$ may 
also be considered as a $TG$--equivariant and linear function $\hf$ on 
$TP$ with values on $\lg$. 
\begin{eqnarray}\label{eqn2}
        \hf : TP &\rightarrow& \lg\;,\nonumber\\
        X_p &\mapsto& \hf(X_p)\quad\mbox{with}\quad 
        \hf(X_p \cdot \xi_g) = {\xi_g}^{-1} \cdot \hf(X_p)\;,
\end{eqnarray}
and $X_p \in TP$, $\xi_g \in TG$. $X_p \cdot \xi_g$  represents the 
action of $TG$ on $TP$ from the right and $\xi_g \cdot \eta$ with $\eta 
\in \lg$ the action of $TG$ on $\lg$ (from the left). It follows 
immediately that this property of $TG$--equivariance of $\hf$ is also 
adapted by the $\omega$. As we shall see, the following equation is 
valid: 
\begin{eqnarray}\label{eqn3}
        \omega(X_p) &=& -\hf(X_p)\;.
\end{eqnarray}
The last topic we would like to discuss, are the gauge tranformations on 
$\omega$ in the light of the above development. It is well known that a 
gauge transformation $\Psi$ can be described by the equivariant function 
$U$ on $P$ with values on $G$. So we have $\Psi(p) = p U(p)$ with 
\begin{eqnarray*}
        U : P &\rightarrow& G
\end{eqnarray*}
and
\begin{eqnarray}\label{eqn4}
        U(p g) &=& g^{-1} U(p) g\;.
\end{eqnarray}
This makes the $G$--action point of view transparent and induces the 
gauge transformation of the connection $\omega$ given by: 
\begin{eqnarray}\label{eqn5}
        (\Psi^* \omega)(X_p) &=& \mbox{Ad}_{{U(p)}^{-1}}  \omega(X_p) + 
        U^{-1} d U (X_p)\;.
\end{eqnarray}
This transformation may also be expressed by the equation 
\begin{eqnarray}\label{eqn6}
        (\Psi^* \omega)(X_p) &=& -(TU X_p)^{-1} \cdot (-\omega(X_p))\;.
\end{eqnarray}
As we see, gauge transformations expressed pointwise are nothing else 
but the $TG$--action on the values of connection. The non--linear term 
in the transformation of a connection may now be explained by the 
$TG$--action on $\lg$. $TG$ is a group of affine type which acts non 
linearly on $\lg$. 

In what follows we discuss in the next section the multiplication rule 
in $TG$, the action of $TG$ on $TP$ and the space $TP$ as a principal 
bundle over $TM$. In section three we present the new definition of a 
connection and focus on its $TG$--equivariance property. In section four 
we consider gauge transformations on a connection from the point of view 
of the $TG$--action and we finally give a summary in section five. 

In the physical literature we found no explicit treatment of the 
$TG$--equi\-variance aspect of connections. The same is valid, to our 
surprise, for the recent mathematical literature we examined. The only 
place we found some, for us very inspiring, but short and compact 
indications on $TG$--equivariance, is the very early paper [7]. Parts of 
section two and three are implicitly or explicitly contained in [7] but 
without connection to the gauge transformations which were the main 
motivation for the present paper. For all these reasons we tried to be 
quite detailed and explicit in our presentations. 

\section{$TP$ as $TG$ principal bundle}
The action of the Lie--group $G$ in the principal bundle $P(M,G)$ is a 
part of its definition. Therefore we expect the $G$--action to be also 
relevant for further structures defined on $P$. This is the case i.e. 
for the connection $\omega$ on $P$ and for the vector valued functions 
on $P$ which correspond to particle fields in the space time $M$.

In this paper we would like to investigate the action of the group $TG$ 
which of course contains $G$ and in addition its Lie algebra $\lg = 
\mbox{Lie}(G)$. For this purpose we have first to determine the 
multiplication law on $TG$ and then the action of $TG$ on $TP$. For the 
reasons explained in the introduction, we shall proceed in a quite 
detailed way. We start with the multiplication law in $TG$, although the 
space $TG$ is isomorphic to the Cartesian product of $\lg$ and $G$, the 
relevant multiplication law is not the so expected trivial one, but it 
makes $TG$ a semi--direct product of $\lg$ and $G$: 
\begin{eqnarray}\label{eqn7}
        TG \times TG    &\rightarrow&   TG\:,\nonumber\\
        (\xi_g, \eta_h) &\mapsto&       \xi_g \cdot \eta_h \ldef TR_h \xi_g 
        + TL_g \eta_h\;.
\end{eqnarray}
The notation $R$ and $L$ is used to the right and left action of $G$ on 
$G$. The inverse of $\xi_g$ is given by 
\begin{eqnarray}\label{eqn8}
        {\xi_g}^{-1}    &=& - TL_{g^{-1}} TR_{g^{-1}} \xi_g\;.
\end{eqnarray}
If we denote with $e$ and $0_e$ the neutral elements in $G$ and $TG$ we 
obtain by explicit calculation $\xi_g \cdot (\xi_g)^{-1} = 0_e$. We have 
of course the injections: 
\begin{eqnarray}\label{eqn9}
\begin{array}{rcl}
        G &\rightarrow& TG\\
        g &\mapsto& 0_g
\end{array}
        &\mbox{and}&
\begin{array}{rcl}
        \lg &\rightarrow& TG\\
        \xi &\mapsto&   \xi_e
\end{array}\quad.
\end{eqnarray}
The verification of associativity is slightly involved and it is 
therefore an advantage to use the isomorphism which contains the 
semi--direct product splitting of $TG$ given by 
\begin{eqnarray}\label{eqn10}
        \alpha : TG     &\rightarrow& \lg \leftsemi G\;,\nonumber\\
        \xi_g   &\mapsto&  (\xi, 0_g) \ldef (TR_{g^{-1}} \xi_g, g)\;.
\end{eqnarray}
Denoting as usual $\mbox{Ad}_G = \mbox{Aut}(\lg)$ with 
\begin{eqnarray}\label{eqn11}
        \mbox{Ad}_g \xi &=& TL_g TR_{g^{-1}} \xi = 0_g \xi 0_{g^{-1}}\;,
\end{eqnarray}
we have for the multiplication rule the multiplication
\begin{eqnarray}\label{eqn12}
        \xi_g \cdot \eta_h &\mapsto& (\xi,g) (\eta,h) = (\xi + \mbox{Ad}_g 
        \eta, g h)\;.
\end{eqnarray}
At this point it should be noticed that of course all the above 
structures are induced naturally by the $T$--functor and the canonical 
isomorphism given in eq. (\ref{eqn10})

In what follows we need several times the action of $TG$ on $\lg$. It 
may be derived from the multiplication law above and the action of $TG$ 
on the $G$--orbit space $TG/G$ given by
\begin{eqnarray}\label{eqn13}
        TG/G &=& \{\xi_g G = TR_h \xi_g | h \in G\}
\end{eqnarray}
which we also denote as $\lg = TG/G$ by misuse of notation. The rule 
of addition in $TG/G$ may be defined by 
\begin{eqnarray}\label{eqn14}
        \xi_g G + \eta_g G &\ldef& (\xi_g + \eta_g) G\:.
\end{eqnarray}
The $TG$--action on $\lg = TG/G$ is given by\footnote{The restriction to 
the subgroup $G$ of $TG$ leads as expected to the adjoined action of $G$
on $\cal G$:
\begin{eqnarray}
	G \times \cal G &\rightarrow& \cal G\nonumber\\
	(g, \eta ) &\mapsto& TR_{g^{-1}}(0_g \cdot \eta) 
	= TR_{g^{-1}} TL_g \eta = \mbox{Ad}_g \eta\nonumber
\end{eqnarray}}
\begin{eqnarray}\label{eqn15}
        TG \times \lg &\rightarrow& \lg\;,\nonumber\\
        (\eta_h, \xi) &\mapsto&  \eta_h \cdot \xi = \mbox{Ad}_h \xi + 
        TR_{h^{-1}} \eta_h\;.
\end{eqnarray}
The action of $TG$ on $TP$ is given canonically from the lifting of the 
$G$--action on $P$ to the tangential space ($T$--functor). So we have in 
an obvious notation from 
\begin{eqnarray}\label{eqn16}
        R : P \times G &\rightarrow& P\;,\nonumber\\
        (p, g) &\mapsto& p g = R_g p = R_p g\:,
\end{eqnarray}
the $TG$--action on $TP$: 
\begin{eqnarray}\label{eqn17}
        TP \times TG &\rightarrow& TP\;,\nonumber\\
        (X_p,\xi_g) &\mapsto& X_p \cdot \xi_g \ldef TR_g X_p + TR_p \xi_g\;.
\end{eqnarray}
In order to facilitate the reading, we give some further notations and some 
examples. For the Maurer--Cartan form $C$ on $G$ given by 
\begin{eqnarray}\label{eqn18}
        C(\xi_g) &=& TL_{g^{-1}} \xi_g\;.
\end{eqnarray}
we often use the expression $\xi(g) = C(\xi_g) \in \lg$. We denote a 
fundamental vector field on $P$ with $\hat{\xi}$ where we have $\xi \in 
\lg$ and  $\hat{\xi}_p = TR_p \xi$.  

We now give some useful examples of the $TG$ action on $TP$. 
\begin{eqnarray}\label{eqn19}
        X_p     \cdot   0_e     &=&     X_p\;,\nonumber\\
        X_p     \cdot   \xi     &=&     X_p +   \hat{\xi}_p\;,\nonumber\\
        X_p     \cdot   0_g     &=&     TR_g X_p\;,\nonumber\\
        0_p     \cdot   \xi     &=&     TR_p \xi \rdef  \hat{\xi}_p\;,
                                        \nonumber\\
        0_p     \cdot   \xi_g   &=&     TR_{p g} C(\xi_g) \rdef TR_{p g}
        \xi(g)\;.
\end{eqnarray}
After this preparation, everything is fixed and as a first result of the 
$TG$--action we may formulate the follwing proposition:\\
{\it The space $TP$ is a principal bundle $TP(TM,TG)$ which structure 
group $TG$ basis manifold $TM$ and canonical projection $T\pi : TP 
\rightarrow TM$. $\pi$ is the canonical projetion in $P(M,G)$, $\pi : P 
\rightarrow M$.}\\
This can be seen from the fact that
\begin{deflist}{(\romannumeral2)}
\item[\romannumeral1)] $TG$ acts freely on $TP$.
\item[\romannumeral2)] $T\pi$ commutes with the group action $TG$.
\end{deflist}
For i) we have to show that with $X_p \cdot \xi_g = X_p$,
\begin{eqnarray}\label{eqn20}
        \xi_g &=& 0_e
\end{eqnarray}
must be valid.\\
From 
\begin{eqnarray}\label{eqn21}
        X_p \cdot \xi_g &=& TR_g X_p + TR_p \xi_g = X_p
\end{eqnarray}
we must first have $g = e$  so that using eq. (\ref{eqn19}), we obtain 
\begin{eqnarray}\label{eqn22}
        X_p \cdot \xi_e &=& X_p +  \hat{\xi}_p =  
        X_p\quad\mbox{and}\quad   \hat{\xi}_p 
        = 0_p 
\end{eqnarray}
which shows that 
\begin{eqnarray}\label{eqn23}
        \xi = 0 \quad\mbox{and}\quad \xi_g = 0_e\;.
\end{eqnarray}
For ii) we have to show that 
\begin{eqnarray}\label{eqn24}
        T\pi X_p &=& T\pi (X_p \cdot \xi_g)
\end{eqnarray}                       
is valid. Using eq. (\ref{eqn19}) and
\begin{eqnarray}\label{eqn25}
        X_p \cdot \xi_g &=& X_p \cdot 0_g + 0_p \cdot \xi_g
\end{eqnarray}
we obtain 
\begin{eqnarray}\label{eqn26}
        X_p \cdot \xi_g &=& TR_g X_p + TR_{pg} \xi(g)\;.
\end{eqnarray}
$TR_{pg}  \xi(g)$ is vertical and 
\begin{eqnarray}\label{eqn27}
        T\pi TR_g X_p &=& T\pi X_p\;,
\end{eqnarray}
so that ii) is indeed valid. 

\section{The $TG$--equivariance of connection}
Having in mind the $G$--equivariance property of connections and the 
$TG$--action on $TP$, as discussed in the previous section, we  
study the space $TP/G$ and ask ourselves what its 
connection with the connections on $P(M,G)$ is. This leads us to a new 
definition of a connection which is of course, as we shall show, 
equivalent to the usual one. The original definition of connections is 
closely related to the $G$--action aspect whereas the definition in this 
section will reveal the $TG$--action aspect and bring to light the most 
characteristic property of connections: the $TG$--equivariance. 

For that purpose we shall first discuss the $G$ orbit space 
\begin{eqnarray}\label{eqn28}
	TP/G    &=& \{ X_p G = \{ TR_g X_p | g \in G\} | p \in P \}
\end{eqnarray}
and we shall show that it is isomorph to the $TP$ associated fibre 
bundle with typical fiber the space $\lg$: 
\begin{eqnarray}\label{eqn29}
        TP/G     &\simeq&      TP \times_{TG} \lg\;.
\end{eqnarray}
The $TG$--action on $\lg$ is given in eq. (\ref{eqn15}). This action is 
non-linear, therefore we do not expect $TP \times_{TG} \lg$ to be a 
vector bundle although it may be represented as $TP \times_{TG} \lg =  
TM \tilde{\times} \lg$. 

The isomorphism in eq. (\ref{eqn29}) follows from the fact that $TP$ is 
a principal bundle and that $G$ is a subgroup of its structure group 
$TG$: 
\begin{eqnarray}\label{eqn31}
  TP/G   &\rightarrow&   TP \times_{TG} (TG/G)\;,\nonumber\\~
  [X_p \cdot \xi_g]_G  &\mapsto& [X_p, \xi_g G]_{TG}\;.
\end{eqnarray}
If we take $\xi_g = 0_e$, this may be seen more perspeciously:
\begin{eqnarray}\label{eqn32}
	X_p G  &\mapsto&   [X_p, 0_e G]_{TG}\;.
\end{eqnarray}
In the fibre bundle $TP/G$, we have to define an addition: for $\pi(p) = 
\pi(q)$ and $p = q g$, we have:
\begin{eqnarray}\label{eqn33}
	X_p G + Y_q G &\ldef& X_p G + (TR_g Y_q) G = (X_p+TR_g Y_q) G\;.
\end{eqnarray}
In what follows we shall give three new definitions of a connection. 
These are closely related to the horizontal lifting of the (original) 
definition of connection:\\
{\it A connection on a principal bundle $P(M,G)$ is a linear section in 
the associated fibre bundle $TP \times_{TG} \lg$ ($\lg = \mbox{Lie}(G)$). 

The action of $TG$ on $\lg$ is given in eq. (\ref{eqn15}). If we denote 
by $\hs$ this section, we have:
\begin{eqnarray}\label{eqn34}
	\hs : TM &\rightarrow& TP/G\;,\nonumber\\
	   V_x &\mapsto&  \hs(V_x)\;.
\end{eqnarray}}
The existence of such a section leads to an equivalent but more abstract 
definition:\\
{\it A connection on a principal bundle $P(M,G)$ is a reduction of the 
structure  group $TG$ of the principal bundle $TP(TM,TG)$ to $G$.}

The most transparent definition we obtain if we consider the above section 
$\hs$ in $TP/G$ as a $\lg$ valued linear function $\hf$ on $TP$:\\
{\it A connection on a principal bundle $P(M,G)$ is a linear 
$TG$--equivariant function on $TP$ with values in $\lg = 
\mbox{Lie}(G)$. $\hf$ is given by 
\begin{eqnarray}\label{eqn35}
	\hf : TP &\rightarrow& \lg\;,\nonumber\\
	  X_p  &\mapsto&   \hf(X_p)\qquad \mbox{with}\nonumber\\
	  \hf(X_p \cdot \xi_g)  & = & (\xi_g)^{-1} \cdot \hf(X_p)\;.
\end{eqnarray}}
At this point it is interesting to note that the function $\hf$ is, as 
we shall show, directly related to the original definition of a 
connection $\omega$ (as a linear function on $TP$) whereas the section 
$\hs$ is directly related to the horizontal lift $\hl$ 
which stems from the $\omega$. The relation between $\hs$ and $\hf$ is 
given by: 
\begin{eqnarray}\label{eqn36}
	\hs(T\pi X_p) & = & [X_p, \hf(X_p)]_{TG}\;.
\end{eqnarray}
The linearity of $\hf$ is related as expected to the linearity of $\hs$: 
\begin{eqnarray}\label{eqn37}
	[X_p,\xi]_{TG} + [Y_p,\eta]_{TG}
& = &   [X_p + \hat{\xi}_p,0]_{TG} + [Y_p + \hat{\eta}_p, 0]_{TG}\;,\nonumber\\
& = &   [X_p \cdot \xi, 0]_{TG} + [Y_p \cdot \eta, 0]_{TG}\;,\nonumber\\
& = &	(X_p + \hat{\xi}_p)G + (Y_p + \hat{\eta}_p)G\;,\nonumber\\
& = &	(X_p + \hat{\xi}_p + Y_p + \hat{\eta}_p)G\;,\nonumber\\
& = &   [X_p + Y_p +  \hat{\xi}_p + \hat{\eta}_p, 0]_{TG}\;,\nonumber\\
& = &   [X_p + Y_p, \xi + \eta]_{TG}
\end{eqnarray}
where we have used the isomorphism of eq. (\ref{eqn32}) and the 
definition of the eq. (\ref{eqn33}).

This corresponds to 
\begin{eqnarray}\label{eqn38}
	[X_p, \hf(X_p)]_{TG} + [Y_p, \hf(Y_p)]_{TG} &=&
	[X_p + Y_p , \hf(X_p) + \hf(Y_p)]_{TG}\;.
\end{eqnarray}
So we obain from the linearity of $\hs$
\begin{eqnarray}\label{eqn39}
	\hs(T\pi X_p) + \hs(T\pi Y_p) = \hs(T\pi (X_p + Y_p))
\end{eqnarray}
the linearity of $\hf$:
\begin{eqnarray}\label{eqn40}
	[X_p,\hf(X_p)]_{TG} + [Y_p, \hf(X_p)]_{TG} &=& 
	[X_p +Y_p, \hf(X_p +Y_p)]_{TG}\nonumber\\
        \hf(X_p + Y_p) &=&  \hf(X_p) + \hf(Y_p)\;.
\end{eqnarray}

For the relation between the horizontal lift $\hl$ and $\hs$, taking 
$\pi(p) = x$ and $T\pi 
X_p = V_x$, we have:
\begin{eqnarray}\label{eqn41}
	\hl_p :   T_xM &\rightarrow& T_pP\;,\nonumber\\
                  V_x  &\mapsto& \hl_p(V_x)\;.
\end{eqnarray}
Using the projection:
\begin{eqnarray}\label{eqn42}
        \mu :    TP    &\rightarrow& TP/G\;,\nonumber\\
                 X_p   &\mapsto& X_p G\;,
\end{eqnarray}
we have
\begin{eqnarray}\label{eqn43}
	\hs_x      &=& \mu \circ \hl_p\;.
\end{eqnarray}
The relation of $\hf$ to connection $\omega$ is given by 
\begin{eqnarray}\label{eqn44}
        \omega(X_p)  &=& -\hf(X_p)\;.
\end{eqnarray}
All we have to show is that the so--defined $\omega = \omega(\hf)$ fulfils 
the original axioms of a connection:
\begin{deflist}{(\romannumeral2)}
\item[\romannumeral1)] $\omega$ is $G$--equivariant.
\item[\romannumeral2)] the restriction of $\omega$ to the fundamental 
	vector fields on $P$ corresponds, after the identification of 
	the fundamental vector fields with the Lie algebra $\lg$, to the 
	identity. 
\end{deflist}
The condition \romannumeral1) is clearly valid since the $\omega$ given 
by $\hf$ is $TG$--equivariant, $G$ is a subgroup of $TG$ and acts on 
$\cal G$ like $\mbox{Ad}_G$.

The condition \romannumeral2) follows from the fact that for  
$\hat{\xi}_p = TR_p \xi$ with $\xi \in \lg$ 
\begin{eqnarray}\label{eqn45}
	\omega(\hat{\xi}_p) &=& 
	-\hf(\hat{\xi}_p)  =  -\hf(0_p \cdot \xi_e)\;,\nonumber\\
  &=&   -{\xi_e}^{-1} \cdot \hf(0_p)\;,\nonumber\\
  &=&   -\xi^{-1} \cdot 0\;,\nonumber\\
  &=&   -(-\xi) = \xi
\end{eqnarray}
where we have used the $TG$--equivariance of $\hf$ (eq. (\ref{eqn35})) and 
the linearity of $\hf$ (eq. (\ref{eqn40})) to evaluate $\hf(0_p)$.

The above considerations showed the equivalence of all the notions of 
connection represented by $\omega$, $\hl$, $\hs$, $\hf$ in every 
combination and particularly in ($\omega$, $\hl$) with ($\hf$, $\hs$). 

At this point we give, for those who feel uneasy with the use of the 
fibre bundle $(TP/G \rightarrow TM) = TP \times_{TG} \cal G$ in the 
definition of a connection in eq. (\ref{eqn34}), a further definition in 
terms of vector bundles. For that purpose we observe that $TP/G 
\rightarrow M$ is a vector bundle with typical fiber $\Reell^m \times 
\cal G$ and $m = \mbox{dim}(M)$. So, denoting this vector bundle by $E = 
(TP/G \rightarrow M)$, we have $E = M \tilde{\times}(\Reell^m \times 
\cal G)$.

This leads to the definition of a connection as a special vector bundle 
homomorphism between $TM$ and $E$:\\
{\it A connection on a principal bundle $P(M,G)$ is a vector bundle 
homomorphism $\hv$:
\begin{eqnarray}
\hv: TM &\rightarrow& E = (TP/G \rightarrow M)
\end{eqnarray}
with the restriction $\nu \circ \hv = \mbox{id}_{TM}$ and $\nu = T\pi 
\circ \mu^{-1}$ the projection $\nu : TP/G \rightarrow TM$. In the case 
of a trivial $P(M,G)$, this definition corresponds to the graph of a 
gauge potential $A$:
\begin{eqnarray}
A : TM &\rightarrow& \cal G
\end{eqnarray}
as it is used in particle physics phenomenology.}
 
\section{$TG$--equivariance and gauge transformations}
In this section we deal with the implications for gauge transformations 
from the $TG$--action point of view. In this way we obtain a new 
formulation for the well known expression for gauge transformations of 
connections in terms of $TG$--elements instead of $G$--elements. 

We consider a gauge transformation as an automorphism on $P(M,G)$ which  
may be expressed equivalently by a function on $P$ with values on $G$ 
[8]:
\begin{eqnarray}\label{eqn46}
	\Psi : P &\rightarrow&  P\;,\nonumber\\
               p &\mapsto& \Psi(p)  = p U(p)
\end{eqnarray}
and 
\begin{eqnarray*}
	U : P &\rightarrow&  G\;,\nonumber\\
            p &\mapsto&  U(p)
\end{eqnarray*}
with
\begin{eqnarray}\label{eqn47}
	U(p g)   &=&       g^{-1} U(p) g\;.
\end{eqnarray}
The automorphism $\Psi$ induces in the space of connections the 
transformation $\Psi^*$ and we have for the connection $\omega$ 
pointwise the expression 
\begin{eqnarray}\label{eqn48}
	(\Psi^* \omega)(X_p) = \mbox{Ad}_{U(p)^{-1}} \omega(X_p) + 
	(U^{*} C)(X_p)
\end{eqnarray}
with $\xi_p \in TP$ and where $C$ is the Maurer--Cartan form on $G$: 
\begin{eqnarray}\label{eqn49}
	C : TG &\rightarrow&  \lg\;,\nonumber\\
        \xi_g   &\mapsto&  C(\xi_g) = TL_{g^{-1}} \xi_g\;.
\end{eqnarray}
We would like now to reexpress the right hand side of eq. (\ref{eqn48}) 
in terms of the $TG$--action on $TP$ and $\lg$. For the action of $TG$ 
on $\lg$ we obtain from 
\begin{eqnarray*}
	\xi_g  \cdot \eta &=& \mbox{Ad}_g \eta + TR_{g^{-1}} \xi_g\;,
\end{eqnarray*}
given in eq. (\ref{eqn15}), the $(\xi_g)^{-1}$ action on $\eta$:
\begin{eqnarray}\label{eqn50}
	(\xi_g)^{-1} \cdot \eta  &=& 
	\mbox{Ad}_{g^{-1}} \eta - TL_{g^{-1}} \xi_g
\end{eqnarray}
or
\begin{eqnarray}\label{eqn51}
	(\xi_g)^{-1} \cdot \eta &=& \mbox{Ad}_{g^{-1}} \eta - C(\xi_g)\;.
\end{eqnarray}
For the $\Psi^* \omega$, expressed in $\hf$, we have pointwise:
\begin{eqnarray}\label{eqn52}
	(\Psi^* \omega)(X_p) &=& \omega(T\Psi X_p) = -\hf(T\Psi X_p)\;.
\end{eqnarray}
Using the equivariance property of $\hf$ (eq. (\ref{eqn35})) and eq. 
(\ref{eqn51}), we obtain: 
\begin{eqnarray}\label{eqn53}
        \hf(T\Psi X_p) &=& \hf(X_p \cdot TU X_p) =
        	(TU X_p)^{-1} \cdot  \hf(X_p)
\end{eqnarray}
and
\begin{eqnarray}\label{eqn54}
        (TU X_p)^{-1} \cdot \hf(X_p) &=&
        	\mbox{Ad}_{U(p)^{-1}} \hf(X_p) - C(TU X_p)\;.
\end{eqnarray}
Eqs. (\ref{eqn52} - \ref{eqn54}) lead to 
\begin{eqnarray}\label{eqn55}
	(\Psi^* \omega)(X_p) &=& 
		-(TU X_p)^{-1} \cdot (-\omega(X_p))
\end{eqnarray}
or equivalently to 
\begin{eqnarray}\label{eqn56}
	(\Psi^* \omega)(X_p) &=& 
	\mbox{Ad}_{U(p)^{-1}} \omega(X_p) + U^* C(X_p)\;.
\end{eqnarray}

On the right hande side of eq. (\ref{eqn55}), the minus in front of 
$\omega(X_p)$ cannot be eliminated because the $TG$--action is not 
linear. Notice also that we obtained in this way eq. (\ref{eqn48}) in a 
very straightforward manner. 

\section{Summary}
The aim of this paper was to set some new lights on several aspects of 
the notion of a connection on a principal bundle $P(M,G)$ and especially 
on its gauge transformations. We expect this to be very useful in 
physical applications.

We proceeded essentially by exploring the properties of the 
$T$--functor, considering the action of the tangential group $TG$ on 
$\cal G = \mbox{Lie}(G)$ and $TP$. It was perceived that $TP$ is also a 
principal bundle $TP(TM,TG)$ and that the fibre bundle $TP/G \rightarrow 
TM$ associated with it ($TP \times_{TG} \cal G$) is useful for a new 
definition of a connection as a linear section on it.

This definition revealed the most characteristic property of a 
connection, the $TG$--equivariance:
\begin{eqnarray}
\omega(X_p \cdot \xi_g) &=& - (\xi_g)^{-1} \cdot (- \omega(X_p))
\end{eqnarray}
where we have used the symbol $\;\cdot\;$ for the $TG$--action on $TP$ 
and $\cal G$.

The $TG$--equivariance property of the connection also gives the most 
plausible explanation for the specific form the expression for gauge 
transformations takes. In particular, the nonlinearity in those 
transformations stems from the nonlinear action of the affine group 
$TG$.
\section*{Acknowledgements}
We would like to thank S. Klaus and M. Kreck for very helpful 
discussions and careful reading of the manuscript. We also thank R. 
Coquereaux for an interesting discussion, and R. Buchert and F. Scheck 
for useful comments.
\section*{References}
\begin{deflist}{8.\qquad}

\item[1.] S. Kobayaschi and K. Nomizu, Foundations of differential geometry,
          Volume 1, John Wiley \& Sons: New York, London, Sydney, 1963.

\item[2.] Y. Choquet--Bruhat, C. De Witt--Morette and M. Dillard--Bleik, 
	  Analysis, Manifolds, and Physics, North--Holland: Amsterdam, 
	  Oxford, New York, Tokyo, 1987.
\item[3.] --~ A. Heil, A. Kersch, N. A. Papadopoulos, B. Reifenh\"auser and 
	  F. Scheck, Structure of the Space of Reducible Connections for 
	  Yang--Mills Theories, Journal of Geometry and Physics: Vol, 7, 
	  n. 4, 1990, 489.
	  --~ A. Heil, A. Kersch, N.A. Papadopoulos and B. Reifenh\"auser, 
	  Abelian Charges in Nonabelian Yang--Mills Theory from the 
	  Stratification of the Space of Gauge Potentials, Annals of 
	  Physics, Vol. 217, No. 2 1992. 
\item[4.] --~ A. Heil, A. Kersch, N.A. Papadopoulos, B. Reifenh\"auser, 
	  F. Scheck and H. Vogel, Anomalies from the Point of View of 
	  G--Theory, Journal of Geometry and Physics: Vol. 6, n. 2, 1989, 
	  237.\\ 
          --~ A. Heil, A. Kersch, N.A. Papadopoulos, E. Reifenh\"auser 
          and F. Scheck: Anomalies from Nonfree Action of the Gauge 
          Group, Annals of Physics: 200, 1990, 206. 
\item[5.] N.A. Papadopoulos, The Role of Stratification of Anomalies, in 
	  J. Debrus and A.C. Hirshfeld (Eds.), Geometry and Theoretical 
	  Physics, Springer-Verlag: Berlin, Heidelberg, 1991, 210.
\item[6.] R. H\"au{\ss}ling, N.A. Papadopoulos and F. Scheck, 
	  Supersymmetry in the Standard Model of Electroweak 
	  Interactions, Physics Letters B303, 1993, 265.
\item[7.] S. Kobayaschi, Theory of connection, Annali di Matematica, 43, 
	  1957, 199.
\item[8.] M. G\"ockeler and T. Sch\"ucker, Differential Geometry, Gauge 
	  Theories, and Gravity, Cambridge University Press: Cambridge, 
	  1987.
\end{deflist}
\end{document}